\documentclass[5p]{elsarticle}
\usepackage{amssymb,amsmath}
\usepackage{lineno}
\usepackage{natbib}
\usepackage{color}
\usepackage{url}
\usepackage{multirow}
\usepackage{epstopdf}
\usepackage{graphicx}
\usepackage{subfigure}
\usepackage{caption}
\journal{Nuclear Instruments and Methods in Physics Research A}
\modulolinenumbers[1]
\usepackage[colorlinks=true,linkcolor=black, citecolor=blue, urlcolor=blue, pdfborder={0 0 0}]{hyperref}
\usepackage{enumitem}
\usepackage{siunitx}

\begin{document}

\begin{frontmatter}
\title{Electromagnetic Simulation and Design of a Novel Waveguide RF Wien Filter for Electric Dipole Moment Measurements of Protons and Deuterons}
\author{J. Slim$^1$, R. Gebel$^2$, D. Heberling$^{1,3}$, F. Hinder$^{2,4}$, D. H{\"o}lscher$^1$, A. Lehrach$^{2,3,4}$, B. Lorentz$^{2}$, S. Mey$^{2,4}$, A. Nass$^2$,\\ F. Rathmann$^2$, L. Reifferscheidt$^5$,  H. Soltner$^5$, H. Straatmann$^5$,  F. Trinkel$^{2,4}$,   and J. Wolters$^5$}
\address{
$^1$ Institute of High-Frequency Technology, Rheinisch-Westf{\"a}lische Technische Hochschule Aachen, 52074 Aachen, Germany\\
$^2$ Institute of Nuclear Physics (IKP), Forschungszentrum J{\"u}lich GmbH, 52428 J{\"u}lich, Germany\\
$^3$ JARA-FAME (Forces and Matter Experiments), Forschungszentrum J{\"u}lich and RWTH Aachen University, 52056 Aachen, Germany\\
$^4$ III. Physikalisches Institut B, Rheinisch-Westf{\"a}lische Technische Hochschule Aachen, 52074 Aachen, Germany\\
$^5$ Central Institute of Engineering and Analytics (ZEA), Forschungszentrum J{\"u}lich GmbH, 52428 J{\"u}lich, Germany}

%
%

\begin{abstract}
The conventional Wien filter is a device with orthogonal static magnetic and electric fields, often used for velocity separation of charged particles. Here we describe the  electromagnetic design calculations for a novel waveguide RF Wien filter that will be employed to solely manipulate the spins of protons or deuterons at frequencies of about 0.1 to 2\,MHz at the COoler SYnchrotron COSY at J{\"u}lich. 
The device will be used in a future experiment that aims at measuring the proton and deuteron electric dipole moments, which are expected to be very small. Their determination, however, would have a huge impact on our understanding of the universe.  
\end{abstract}

\end{frontmatter}

\section{Introduction \label{sec:introduction}}
Recently, the use of a Wien filter operated at radio frequency (RF) has been proposed as a tool to  search for electric dipole moments (EDMs) of protons and deuterons in storage rings~\cite{Rathmann:2013rqa,morse}. The occurrence of EDMs of elementary particles is intimately connected to the matter-antimatter asymmetry observed in the universe~\cite{doi:10.1146/annurev.nucl.49.1.35}, which the Standard Model of elementary particle physics fails to describe. A non-zero EDM measurement would point to new physics beyond the Standard Model~\cite{Dekens2014}.
 
A Wien filter provides orthogonal electric and magnetic fields, usually generated by a parallel plate capacitor and encircling coils, such that the Lorentz force for charged particles traveling with a specific velocity orthogonal to both fields vanishes~\cite{salomaa}. This principle has found widespread applications not only in mass spectro\-meters, but also in electron microscopes and ion optics~\cite{wienmicro}. A few reports have been dedicated to its use in accelerator facilities, but most of them describe static Wien filters~\cite{orloff2008handbook,steiner}.  The electric and magnetic fields of an RF Wien filter can be used to manipulate the spins of particles in a storage ring, and, as shown in~\cite{Rathmann:2013rqa,morse}, open up the possibility for a measurement of the EDMs of protons and deuterons.

The objective of the present publication is to describe the design of a novel waveguide RF Wien filter for the search for the EDMs of protons and deuterons at COSY~\cite{Maier19971,PhysRevSTAB.18.020101}, pursued by the JEDI collaboration\footnote{JEDI collaboration \url{http://collaborations.fz-juelich.de/ikp/jedi}}. The fact that proton and deuteron EDMs are expected to be very small calls for precise design and manufacturing. We must admit that the spin-tracking tools required to provide specifications for the design of the RF Wien filter have not yet been fully developed. Recently, spin-tracking studies with an \textit{ideal} RF Wien filter were carried out~\cite{Rosenthal:2015jzr,doi:10.1142/S2010194516600995}, but these do not yet take into account a number of important systematic effects, such as fringe fields, unwanted field components, positioning errors, and non-vanishing Lorentz forces.  Therefore, an approach was adopted here to try to realize the best possible device based on state-of-the-art technologies. 

A prototype RF Wien filter, based on an already existing RF dipole with radial magnetic field ~\cite{PhysRevLett.93.224801,CERN:Courier}, was recently developed and used at COSY.  (The use of RF dipoles and solenoids to manipulate stored polarized beams is discussed in~\cite{lee}.) The RF dipole was equipped with horizontal electric field plates in order to provide an RF Wien filter configuration with vertical electric and horizontal magnetic field~\cite{Mey:2015xbq,doi:10.1142/S2010194516600946}. The RF electromagnetic field was generated using two coupled resonators; one that generates the electric field and another one that generates the magnetic field. The approach of using separate systems to generate electric and magnetic fields, however, neglected the inherent coupling between the electric and magnetic fields. Therefore, the approach described here, is based on a novel waveguide system where by design the orthogonality between electric and magnetic fields is accomplished.

The paper is organized as follows:
\begin{itemize}
\item Section~\ref{sec:design} describes the mechanical design of the  RF Wien filter. The parallel-plates wave\-guide and its mechanical structure are described in Section~\ref{sec:waveguide}, the driving circuit in Section~\ref{sec:driving_circuit}. The results of the electromagnetic field simulations, the Lorentz force compensation, and the optimization of the electric and magnetic field homogeneity by shaping  the electrodes is discussed in Section~\ref{sec:field_homogenity}.

\item Section~\ref{sec:beam_dynamics_simulations} describes the beam dynamics simulations. A Monte Carlo simulation was carried out in order to quantify the unwanted field components of the RF Wien filter with a realistic phase-space distribution of the beam. A comparison of the prototype RF Wien filter~\cite{Mey:2015xbq,doi:10.1142/S2010194516600946} and the wave\-guide RF Wien filter is given in Section~\ref{sec:comparison}.

\item Section~\ref{sec:thermal} presents results of thermal simulations based on the power losses in the different materials of the device.
\end{itemize}
 
\section{Design \label{sec:design}} 
\subsection{Waveguide design \label{sec:waveguide}}
The RF Wien filter shall be operated at frequencies of about 100\,kHz to 2\,MHz. The maximum acceptable length for the vacuum vessel, given by space restrictions at COSY, is 870\,mm, which corresponds to approximately $1/300$ of the length of the electromagnetic wave (at $\SI{1}{MHz}$). The transverse electromagnetic (TEM) mode of a parallel-plate wave\-guide fulfills the requirement of orthogonal electric and magnetic fields. 

In order to maintain an overall vanishing Lorentz-force, and to provide minimal unwanted field components, it is a priori not clear whether a small or a large beam size is advantageous. Within the COSY ring, the PAX low-$\beta$ section~\cite{PhysRevSTAB.18.020101} offers the possibility to vary the beam size by about a factor of three, therefore the RF Wien filter will be installed at this location.

The design calculations for the waveguide RF Wien filter were carried out using a full-wave simulation with CST Microwave Studio\footnote{CST - Computer Simulation Technology AG, Darmstadt, Germany, \url{http://www.cst.com}}, and the electric and magnetic fields were modeled with an accuracy of $10^{-6}$. Because of the  high expectations on field homogeneity, an approach was adopted that allows one to calculate the electric and magnetic fields without additional assumptions, such as quasi-static approximations or the like\footnote{Each simulation required up to 12 hours of computing time on a 4-Tesla C2075 GPU cluster\footnotemark[4],  with 2 six-core Xeon E5 processors\footnotemark[5] and a RAM capacity of 94\,GB.}.
\footnotetext[4]{Nvidia Corporation, Santa Clara, California, USA, \url{http://www.nvidia.com/object/tesla-workstations.html}}\addtocounter{footnote}{+1}  
\footnotetext[5]{Intel Corporation, Santa Clara, California, USA, \url{http://www.intel.com/content/www/us/en/processors/xeon/xeon-processor-e5-family.html}}\addtocounter{footnote}{+1}  

The system comprises a power source, a parallel plate waveguide and a load. The plates are fed on their front and rear sides by an RF current, which distributes on the surfaces, thereby generating the required electromagnetic field in the enclosed space. Deviations from the ideal orthogonality and homogeneity of the fields are due to the finite size of the plates and their limited conductivities.

The intended working frequencies of the RF Wien filter are calculated according to
\begin{equation}
f_{\text{RF}}=f_{\text{rev}} \lvert {k+\gamma G} \rvert, k \in \mathbb{Z}\,
\end{equation}
where $k$ is the harmonic number, $G$ the gyromagnetic anomaly, $\gamma G$ the spin tune~\cite{PhysRevLett.115.094801}, and $f_{\text{rev}}$ the revolution frequency. With respect to the search for the EDMs of deute\-rons and protons, experiments at a number of harmonics shall be carried out for systematic reasons. Table~\ref{tab:harmonic} summarizes the resonance frequencies for deuterons at a momentum of $\SI{970}{MeV/c}$ and for protons at $\SI{520}{MeV/c}$, for which electron-cooled beams with compensated cooler solenoid are available at COSY. As indicated in Table~\ref{tab:harmonic} for deuterons (protons), five (four) harmonics are in the operating frequency range of $\SI{100}{kHz}$ to $\SI{2}{MHz}$. In the following, we restrict the discussion to deuterons at a momentum of $\SI{970}{MeV/c}$.

\begin{table*}[htb]
\renewcommand{\arraystretch}{1.2}
\centering
\caption{Operating frequencies $f_\text{RF}$ of the waveguide RF Wien filter for deuterons ($d$) at a momentum of 970\,MeV/c and for protons ($p$) at $\SI{520}{MeV/c}$ in COSY for the harmonic numbers $k$. The frequencies $f_\text{RF}$ shown in bold fit in the frequency range from $\SI{100}{kHz}$ to 2\,MHz. The revolution frequency $f_\text{rev}$, G factors, Lorentz $\beta$ and $\gamma$, and the spin tune $\gamma G$ are given as well.}
\label{tab:harmonic}
\begin{tabular}{cccccc|ccccccc}\hline
         & &         &            &         &          &       & \multicolumn{5}{c}{$f_\text{RF}$\,[kHz]}\\
 & $f_\text{rev}$\,[kHz] & G          & $\beta$ & $\gamma$ & $\gamma G$ & $\scriptstyle k=-4$ & $\scriptstyle k=-3$&  $\scriptstyle k=-2$      & $\scriptstyle k=-1$     & $\scriptstyle k=0$     & $\scriptstyle k=+1$    & $\scriptstyle k=+2$ \\\hline
$d$      & $750.2$ & $-0.143$   & $0.459$ & $1.126$ & $-0.161$   & $3121.6$ & $2371.4$ &  $\mathbf{1621.2}$  & $\mathbf{871.0}$  & $\mathbf{120.8}$ & $\mathbf{629.4}$ & $\mathbf{1379.6}$\\
$p$      & $791.6$ & $1.793$    & $0.485$ & $1.143$  & $2.050$    & $\mathbf{1543.9}$ & $\mathbf{752.2}$&  $39.4$  & $\mathbf{831.0}$ & $\mathbf{1622.7}$ & $2414.3$ & $3206.0$ \\\hline
\end{tabular}
\end{table*}

The Lorentz force is given by
\begin{equation}
\vec F_\text{L}  = q \left(\vec E + \vec v \times \vec B\right)\,,
\label{eq:lorentz} 
\end{equation}
where $q$ is the charge of the particle, $\vec{v}=c(0,0,\beta)$ is the velocity vector,  $\vec E=(E_x,E_y,E_z)$ and $\vec B=\mu_0(H_x,H_y,H_z)$ denote the components of the electric and magnetic fields, and $\mu_0$ the vacuum permeability. For a vanishing Lorentz force $\vec{F}_\text{L}=0$, the required field quotient $Z_q$ is determined, which yields
\begin{align}
E_x & = -c \cdot \beta \cdot \mu_0 \cdot H_y\,, \nonumber \\
Z_q=-\frac{E_x}{H_y} & = c \cdot \beta \cdot \mu_0 \approx 173 \, \, \Omega\,.
\label{eq:zq}
\end{align}

\begin{figure*}[!]
\centering
\includegraphics[width=11cm]{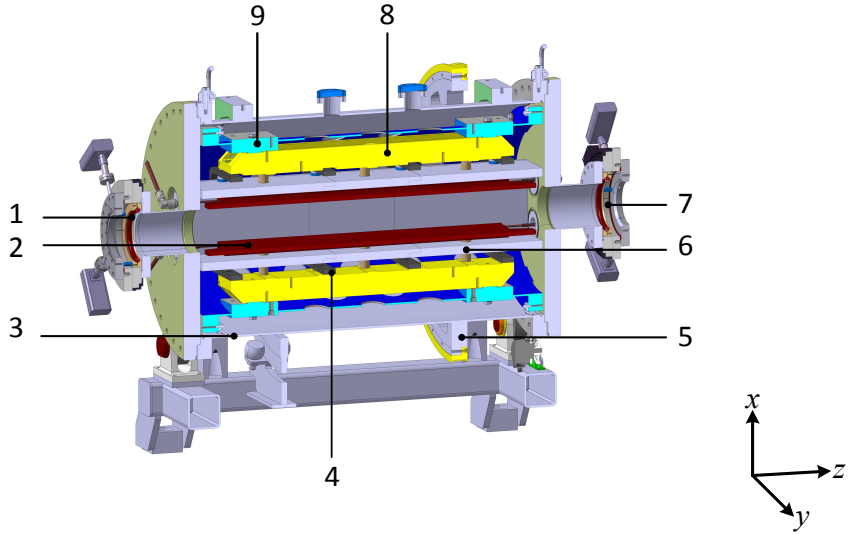}
\caption{\label{fig:structure} Design model of the RF Wien filter showing the parallel-plates waveguide and the support structure. 1: beam position monitor (BPM); 2: copper electrodes; 3: vacuum vessel; 4: clamps to hold the ferrite cage; 5: belt drive for $\ang{90}$ rotation, with a precision of $\ang{0.01}$ ($\SI{0.17}{mrad}$); 6: ferrite cage; 7: CF160 rotatable flange; 8: support structure of the electrodes; 9: inner support tube.  The axis of the waveguide points along the $z$-direction, the plates are separated along $x$, and the plate width extends along $y$. During the EDM studies, the main field component $E_x$ points radially outwards and $H_y$  upwards with respect to the stored beam.}
\end{figure*}

Figure~\ref{fig:structure} shows a cross section of the parallel-plates wave\-guide RF Wien filter. The axis of the waveguide points along the beam direction ($z$). The plates are separated by $\SI{100}{mm}$ along the $x$-direction. The width of the plates is $\SI{182}{mm}$. This setup ensures that during the EDM studies, the main component of the electric field ($E_x$) points radially inwards in $-x$-direction, and the main component of the magnetic field ($H_y$)  upwards in $y$-direction with respect to the stored beam.  The system is placed inside a cylindrical vacuum vessel. The electrodes are surrounded by ferrite blocks made of CMD5005\footnote{National Magnetics Group, Inc., Pennsylvania, USA,  \url{http://www.magneticsgroup.com/m_ferr_nizn.htm}}. The copper electrodes are shaped in order to improve the homogeneity of the electric and magnetic fields and to minimize the Lorentz force, as explained in detail in Section~\ref{sec:EM-simulation}.

\begin{figure*}[!t]
\centering
\includegraphics[width=10cm]{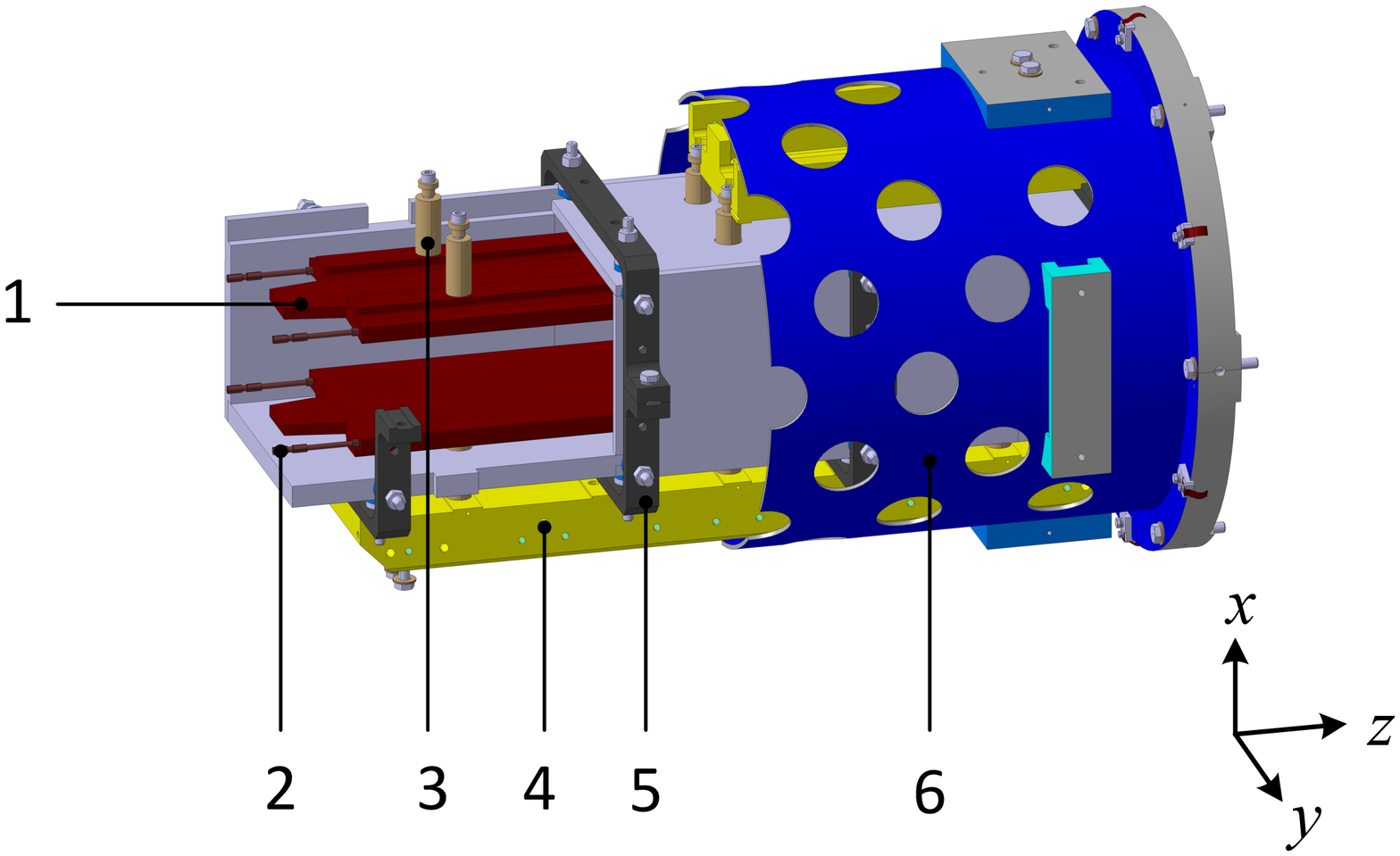}
\caption{\label{fig:support} Inner support structure of the RF Wien Filter. 1: copper electrodes with the trapezium shaping at the edges; 2: specially designed connector; 3: ceramic insulator between the electrodes and the support structure. A stainless-steel screw is located inside to connect the electrodes to the support structure; 4: support of the electrodes; 5: clamps to support the ferrite cage; 6: inner tube support structure.}
\end{figure*}

A sophisticated support structure, shown in Fig.~\ref{fig:support}, was designed to ensure a high degree of rigidity, precision alignment and most importantly, low distortion of the generated electromagnetic field. The electrodes are made from double T-shaped copper plates for better stiffness and stability. They are mounted with 12 stainless-steel screws on the main support structure of the electrodes. An inner stainless-steel tube is used to hold, align and mount the structure with high precision ($\approx 10$\,$\mu$m). 

The angular position of the RF Wien filter with respect to the beam axis can be chosen to allow for a rotation of the field direction by $\ang{90}$ with an angular precision of $\ang{0.01}$ ($\SI{0.175}{mrad}$) using a belt drive (see label 5 in Fig.~\ref{fig:structure}), without breaking the vacuum. This feature is foreseen in order to fine tune the orientation of the RF Wien filter~\cite{doi:10.1142/S2010194516600995}, and also to exchange the role of electric and magnetic fields during the EDM measurements for  investigations of the systematics. 

The parallel-plates waveguide constitutes a transmission line structure, which has been analyzed in detail in~\cite{slimpstp2015}. For the waveguide to be used as a Wien filter, wave mismatch must be introduced into the structure via a reflection, ensured by adding a resistor that controls the reflection coefficient. Thereby, the characteristic field quotient $Z_q$, given in Eq.~(\ref{eq:zq}), can be adjusted to the required value. 

The multi-input feedthroughs ensure a homogeneous current distribution over the electrodes, the two parallel plates are connected to the amplifier and the load via four high-frequency, high-power CF 40 feedthroughs\footnote{VACOM Vakuum Komponenten $\&$ Messtechnik GmbH, Jena, Germany, \url{http://www.vacom.de}}. The electrodes are connected to the feedthroughs via specially designed connectors, and $3$\,dB power splitters are installed to feed each plate at the edges. 

\begin{figure}[hbt]
\centering
\includegraphics[width=0.9\columnwidth]{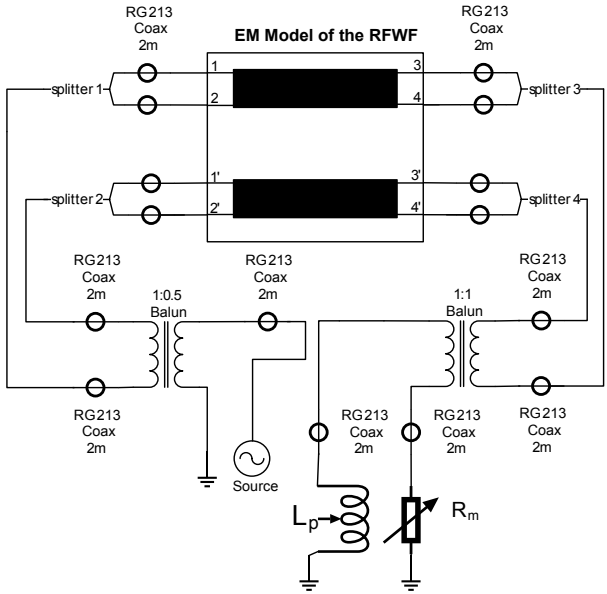}
\caption{\label{fig:circuit} Schematic of the driving circuit for the RF Wien filter. The condition for minimal Lorentz force is met using two adjustable devices, the resistor $R_\text{m}$ and the inductance $L_\text{p}$.}
\end{figure}

The RF Wien filter comprises a cage of ferrite blocks from CMD5005, which contains the electromagnetic fields within the Wien filter and enhances the field homogeneity.  CMD5005 is a NiZn-type high-permeability ferrite material. Ref.~\cite{Hahn_rhicabort} provides data of a dispersive model of the complex permeability $\mu_r$ that have been used in the full-wave simulator at frequencies from $\SI{100}{kHz}$ to 2\,MHz. At a frequency of $\SI{871}{kHz}$, for instance, $\mu_r=1449 + i 467$. The relative permittivity $\epsilon_r$ is around 25~\cite{zhang}. CMD5005 elements come in blocks with a maximum length of 330\,mm. The ferrites are held together by clamps in a symmetric arrangement. To make a connection between the electrodes and the main support structure possible, 12 boreholes are drilled in the ferrites. The screws supporting the electrodes are insulated from the ferrites by ceramic cylindrical insulators. 

On each side of the CF100 entry and exit ports, 4 CF16 feedthroughs are connected for the beam position monitors (BPMs)~\cite{hinder}, especially designed to ensure the alignment of the beam with respect to the axis of the RF Wien filter.

\subsection{Driving circuit \label{sec:driving_circuit}}
\begin{figure}[hbt]
\centering
\includegraphics[width=0.9\columnwidth]{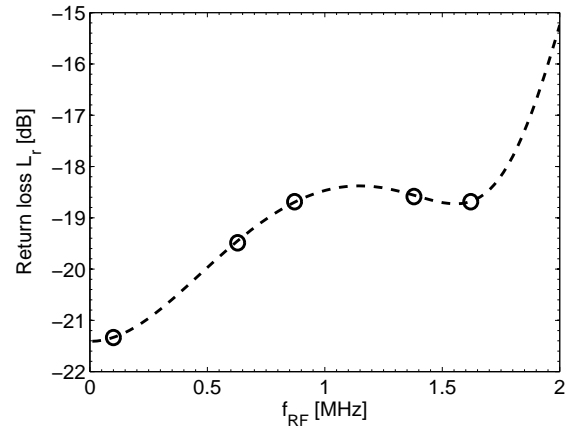}
\caption{\label{fig:s_11} Return loss $L_\text{r}$  of the electromagnetic simulation of the circuit, shown Fig.~\ref{fig:circuit}. For the entire range of spin harmonics (from 0 to $\SI{2}{MHz}$), $L_\text{r}$ is below $\SI{-15}{dB}$. Wideband matching is  achieved with a $1:0.5$ Balun.}
\end{figure}

Ideally, the driving circuit connects a load resistor directly to the electrodes, as shown in~\cite{slimpstp2015}. The amplifier (with 50 $\Omega$ internal impedance) is connected to the Wien filter via a ($1:0.5$) Balun~\cite{hickman} which transforms the impedance and converts the unbalanced output of the amplifier into a balanced one. A schematic of the driving circuit is shown in Fig.~\ref{fig:circuit}. One of the main characteristics of Balun transformers is their broadband response. The return loss $L_\text{r}$ of the driving circuit (Fig.~\ref{fig:circuit}), calculated using CST Design Studio, is depicted in Fig.~\ref{fig:s_11}. 

Another ($1:1$) Balun is used to connect the Wien filter to the load resistor. Introducing the cables into the schematic induces a phase shift between the electric and magnetic field,
consequently, the field quotient $Z_q$, given in Eq.~(\ref{eq:zq}), becomes complex. 
This effect can be compensated with an additional inductance $L_\text{p} = 8\,\mu\text{H}$  (as indicated in Fig.~\ref{fig:circuit}), and an almost purely real-valued field quotient is obtained, \textit{i.e.}, $Z_q = 173\,\angle 0.1 ^\circ \Omega$. 

\subsection{Optimization of field homogeneity \label{sec:field_homogenity}}
\label{sec:EM-simulation}
For minimal Lorentz forces, the electric and magnetic forces must be matched at the center and the edges of the RF Wien filter. As mentioned before, the ratio between the electric and magnetic fields inside the RF Wien filter can be controlled via the load resistor $R_\text{m}$, and the inductance $L_\text{p}$ (see Fig.~\ref{fig:circuit}). A special solution, however, is required at the edges, where the slope of the curves of the electric and magnetic forces are not the same. According to our simulations, a simple parallel-plates waveguide deflects particles passing through the device in the same directions at the exit and entry points. Under these cir\-cum\-stances, the overall Lorentz force \textit{cannot} be zero when the Lorentz force in the center is zero.

\begin{figure}[t]
\centering
\includegraphics[width=0.9\columnwidth]{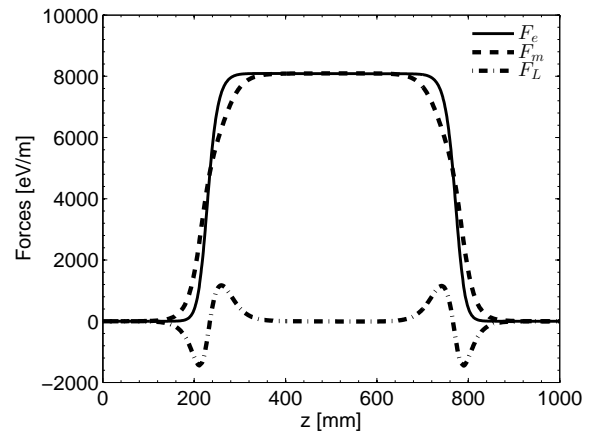}
\caption{Electric force $F_\text{e}$, magnetic force $F_\text{m}$, and Lorentz force $F_\text{L}$  inside the RF Wien filter, for the geometry shown in Fig.~\ref{fig:structure}.   The integral Lorentz force is of the order of $10^{-3}$\,eV/m. The trapezoid-shaped electrodes at the entrance and exit of the RF Wien filter determine the crossing of electric and magnetic forces.}
\label{fig:lorentz}
\end{figure}

Our solution aims at a decomposition of the kicks at each side of the RF Wien filter into two deflections of opposite sign in such a way that  both deflections average out. Decoupling the fields at the edges, keeping the electric field unchanged while manipulating the magnetic field, is accomplished by retaining the plate separation (spacing between the plates), while altering the width of the plates~\cite{pozar}. In doing so, the slope of the curves of the electric field remains constant while the magnetic field crosses the electric field, as shown in Fig.~\ref{fig:lorentz}. This approach results in trapezoid-shaped edges of the parallel plates, where the field crossing can now be optimized via the geometrical properties of the trapezoid-shaped edges. After performing a series of simulations, an optimum trapezoid depth of $\SI{50}{mm}$, with a short base of $\SI{50.5}{mm}$  and a long base of $\SI{100}{mm}$ was found, as shown in Fig.~\ref{fig:support} (label 1).

Integrating and averaging the Lorentz force $\vec F_\text{L}$, given in Eq.~(\ref{eq:lorentz}), along the axis of the RF Wien filter for the geometry shown in Fig.~\ref{fig:structure} at an input power of $\SI{1}{kW}$, yields
\begin{equation}
\frac{q}{\ell}\int_{-\ell/2}^{\ell/2} \left( \begin{array}{ccc}
                                       E_x - c \beta B_y\\
                                       E_y + c \beta B_x\\
                                       E_z
                                      \end{array} \right) dz=
\left(\begin{array}{ccc}
5.97 \times 10^{-3}\\
7.97 \times 10^{-3}\\
1.27 \times 10^{-21}\\
\end{array}\right) \, \text{eV/m}\,,
\label{eq:lorentz-force-value}
\end{equation}
where $\ell=\SI{1550}{mm}$ denotes the active length of the RF Wien filter, defined as the region where the fields are non-zero. The momentum variation in the beam of about $\Delta p / p = 10^{-4}$ translates into a variation of the required field quotient $Z_q$ of the same order of magnitude [see Eq.~(\ref{eq:zq})], and therefore, the resulting values for the Lorentz forces, given in Eq.~(\ref{eq:lorentz-force-value}), are acceptable.

\begin{figure*}[htb]
\centering
 \includegraphics[width=0.45\textwidth]{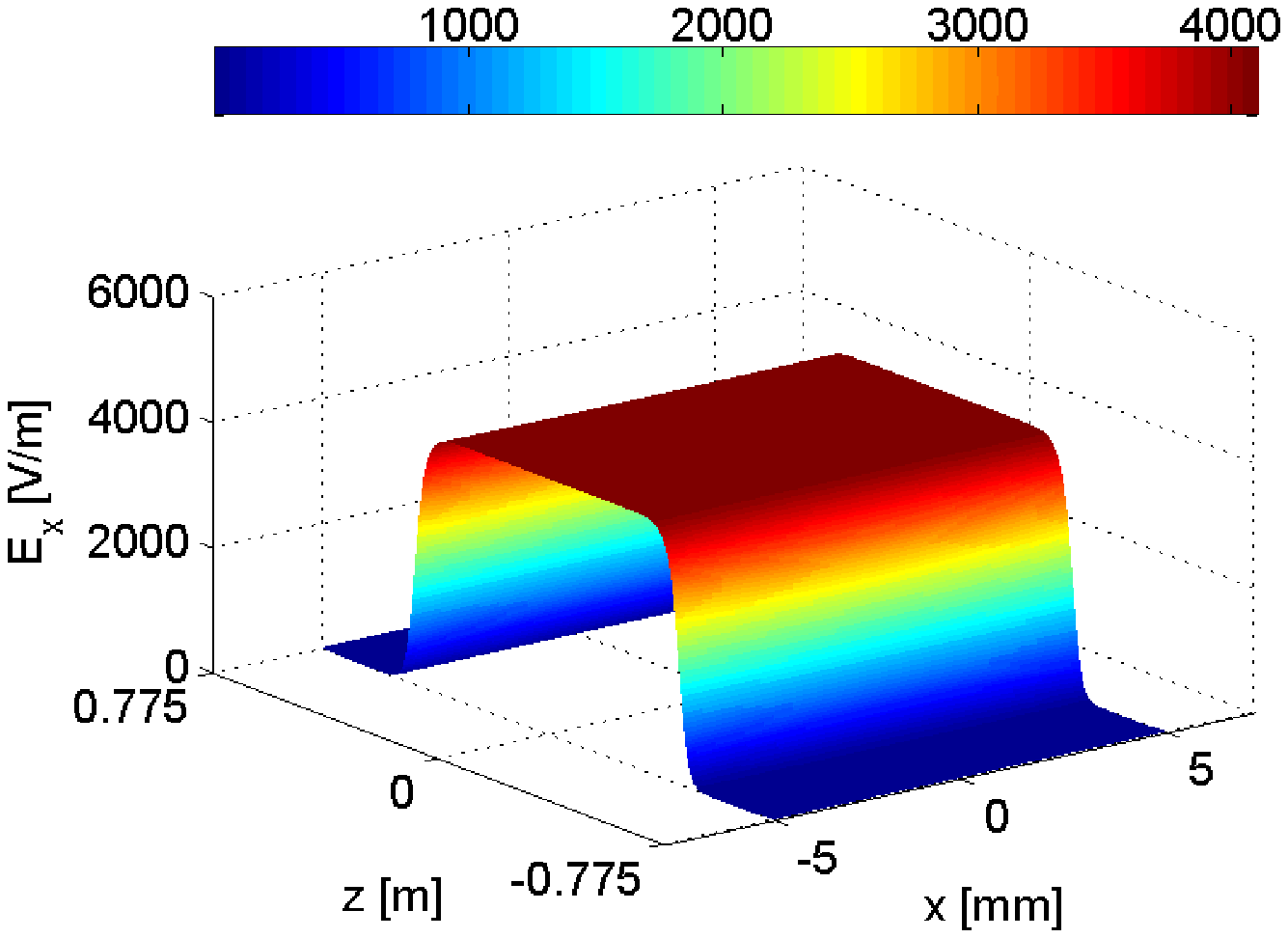}
 \includegraphics[width=0.45\textwidth]{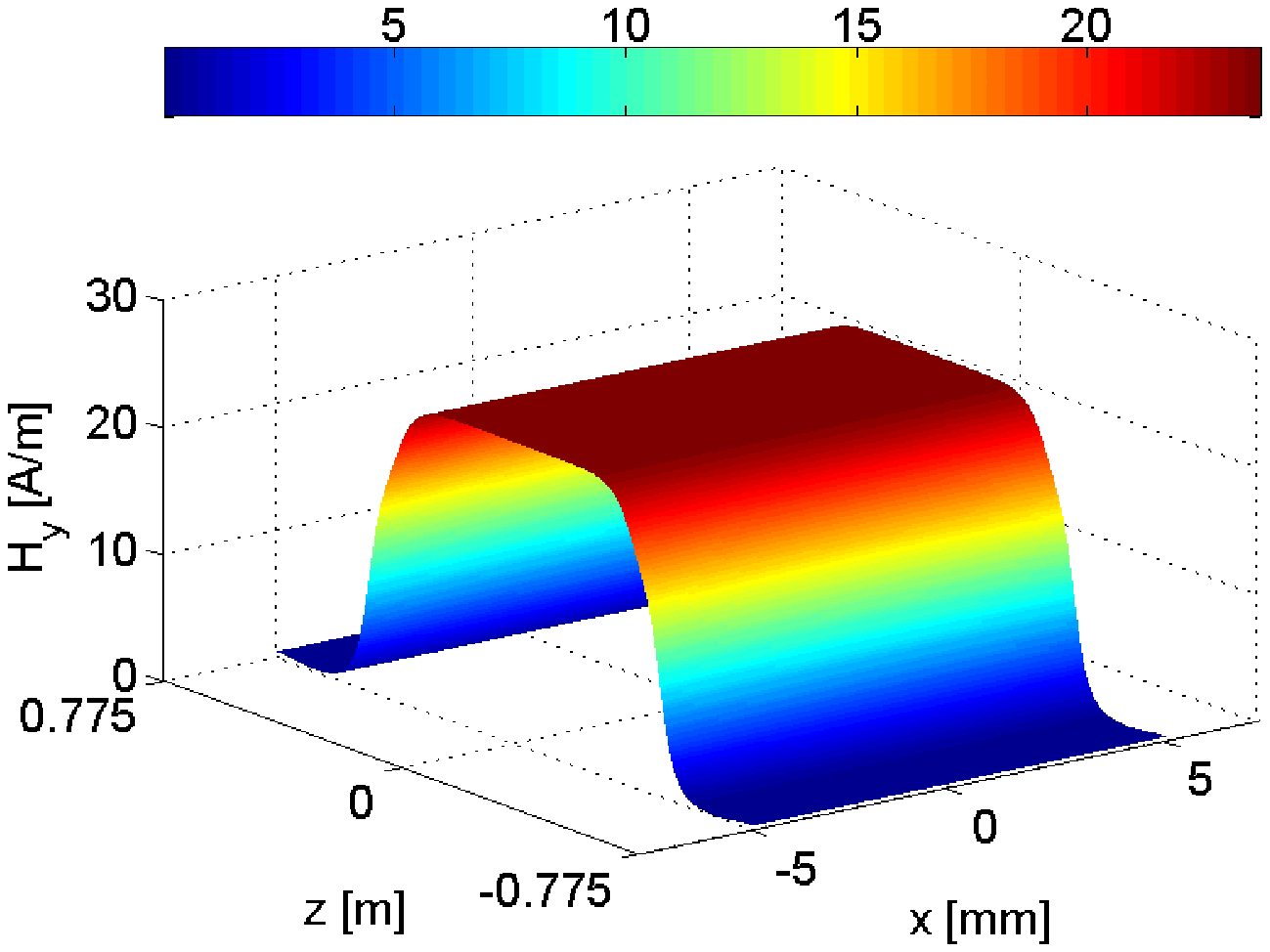}
\caption{\label{fig:main_fields} Main components of the electric field $E_x$ (left panel) and magnetic field $H_y$ (right) of the waveguide RF Wien filter, for the geometry shown in Figs.~\ref{fig:structure} and \ref{fig:support}.}
\end{figure*}

The waveguide RF Wien filter can be rated according to its ability to manipulate the spins of the stored particles, and as a figure of merit, the field integral of $\vec B$ along the beam axis is evaluated, yielding for an input power of $\SI{1}{\,kW}$,
\begin{equation}
 \int_{-\ell/2}^{\ell/2} \vec B dz =  
 \left(\begin{array}{ccc}
  2.73 \times 10^{-9}\\
  2.72 \times 10^{-2}\\
  6.96 \times 10^{-7}\\
 \end{array}
\right)\,\text{T\,mm}\,.
\end{equation}
Under these conditions, the corresponding integrated electric field components are given by
\begin{equation}
 \int_{-\ell/2}^{\ell/2} \vec E dz =  
 \left(\begin{array}{rrr}
  3324.577 \\
  0.018\\
  0.006\\
 \end{array}
\right)\,\text{V}\,.
\end{equation}
Figure~\ref{fig:main_fields} shows the main (wanted) components of the electric and magnetic fields, $E_x$ and $B_y$, in the $xz$ plane.

\begin{figure}[hbt]
\centering
\includegraphics[width=1\columnwidth]{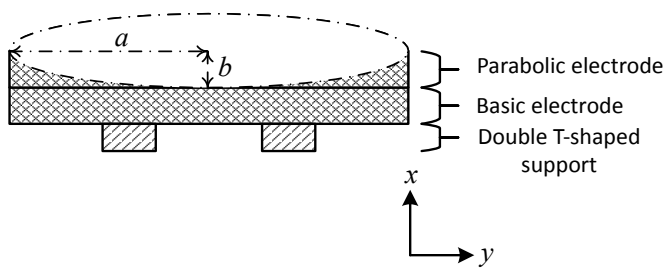}
\caption{\label{fig:parabolic_design} Schematic view of the cross section of the bottom electrode. The double T-shaped support is used to fix the electrode to the support structure. The parabolic shape is constructed using an ellipse with a major radius of $a = \SI{91}{\,mm}$, and a minor radius of $b=\SI{6}{mm}$. Sharp edges are avoided using a rounding radius of $1\,$mm.}
\end{figure}

In order to further increase the field homogeneity, the geometric parameters such as the width and the surface shape of the copper electrodes, and also the geometric parameters of the ferrite blocks and their distance to the metallic support structure, including the surrounding vacu\-um vessel were optimized. The simulations showed that parabolically-shaped electrode surfaces instead of flat ones substantially improve the local homogeneity of the electric field along the RF Wien filter. The parameters of the optimized parabolic electrodes are indicated in Fig.~\ref{fig:parabolic_design}, with a major radius of $\SI{91}{mm}$, and a minor radius of $\SI{6}{mm}$, where the sharp edges have been rounded using a $1$\,mm radius.  In the case of flat electrodes, the electric field varies up to $\SI{8}{\text{V/m}}$, and by parabolically shaping the electrodes, as shown in Fig.~\ref{fig:parabolic_design}, the electric field variation does not exceed $\SI{0.1}{\text{V/m}}$.

\begin{figure*}[htb]
\centering
\subfigure[Flat: $\left|\frac{|E_x|- \left\langle E_x \right\rangle}{|E_x|} \right|$]{\includegraphics[width=0.32\textwidth]{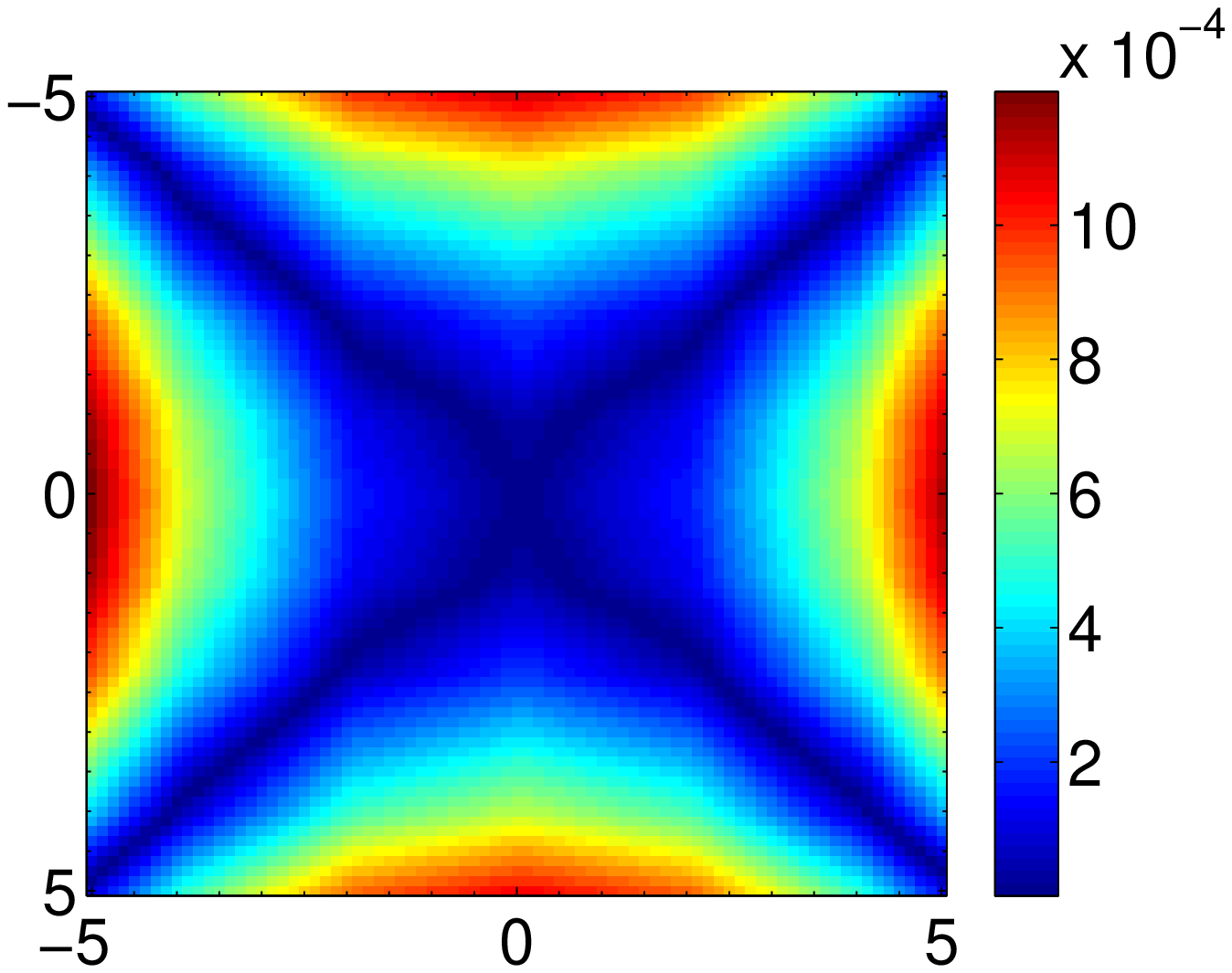}}
\subfigure[Flat: $\left|\frac{|H_y|- \left\langle H_y \right\rangle}{|H_y|} \right|$]{\includegraphics[width=0.32\textwidth]{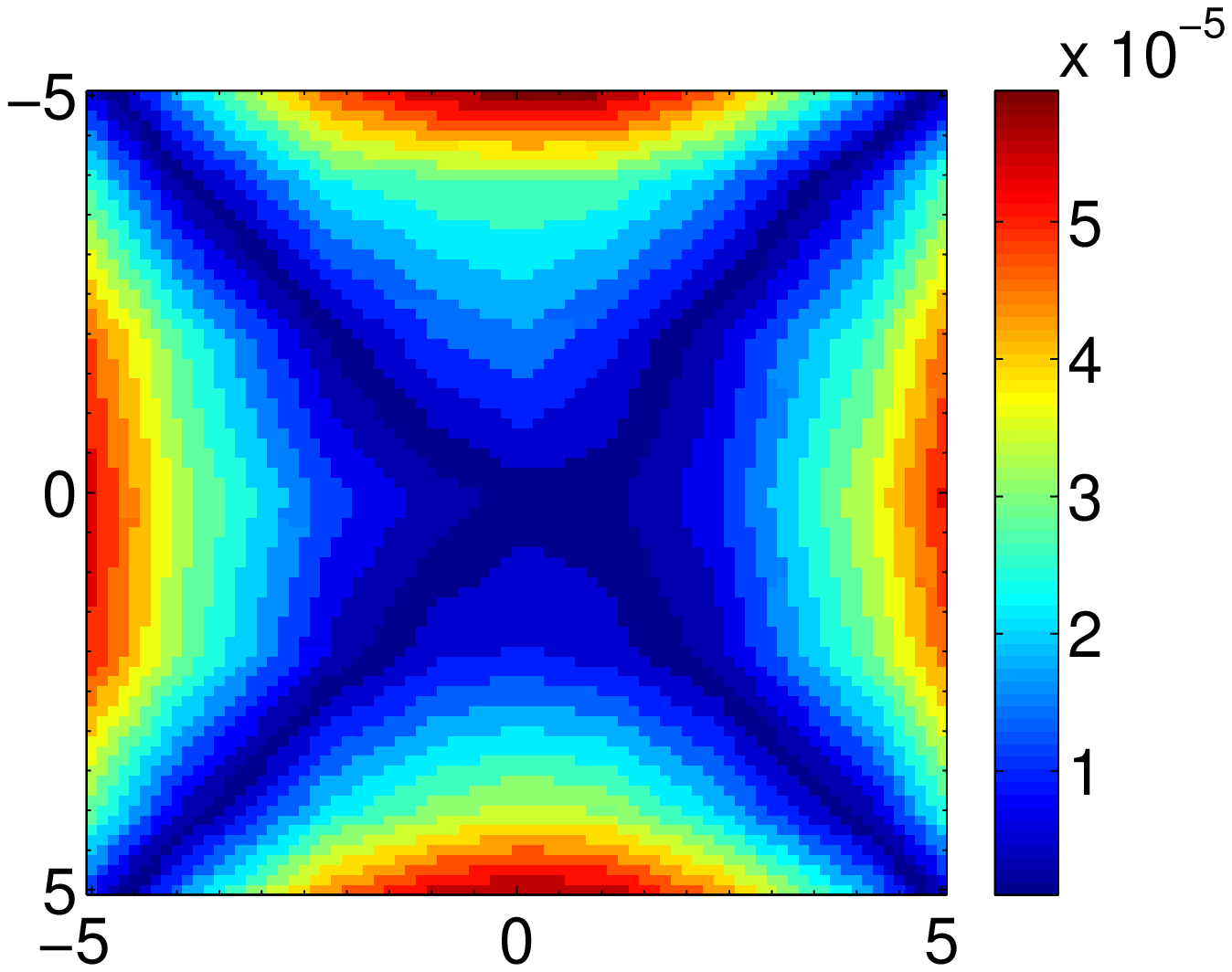}}
\subfigure[Flat: Lorentz force in $\text{eV/m}$]
{\includegraphics[width=0.32\textwidth]{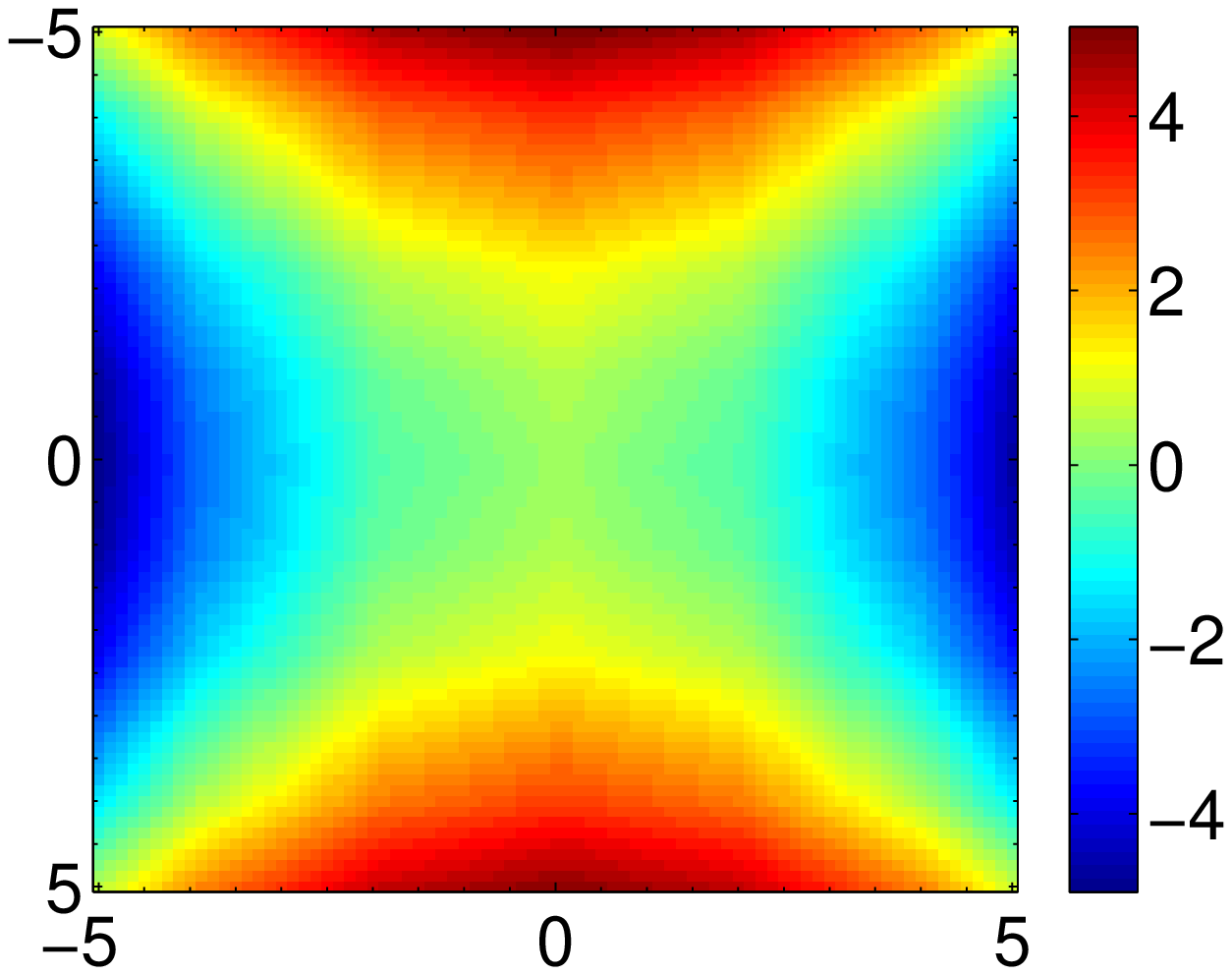}}
\subfigure[Parabolic: $\left|\frac{|E_x|- \left\langle E_x \right\rangle}{|E_x|} \right|$]{\includegraphics[width=0.32\textwidth]{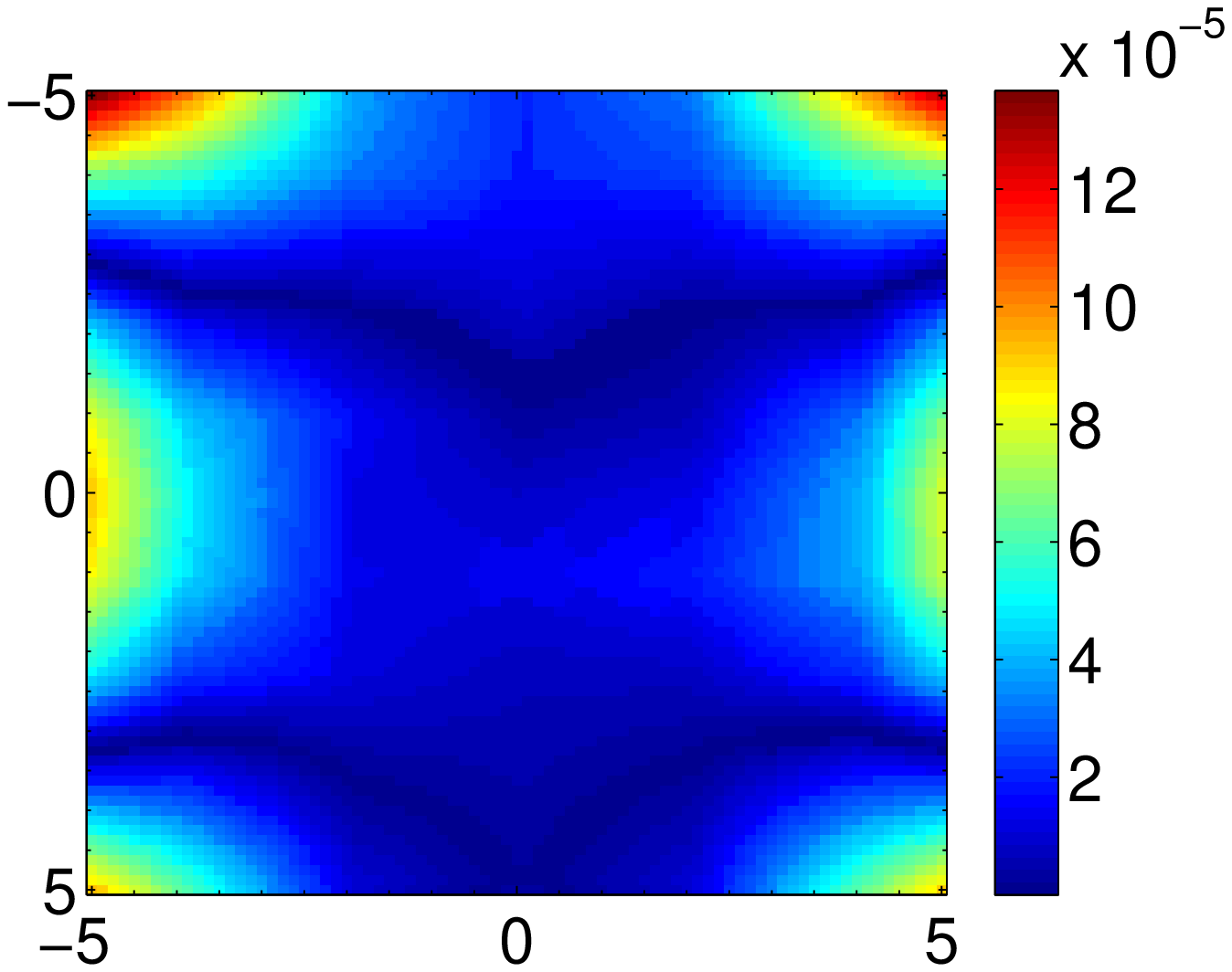}}
\subfigure[Parabolic:  $\left|\frac{|H_y|- \left\langle H_y \right\rangle}{|H_y|} \right|$]{\includegraphics[width=0.32\textwidth]{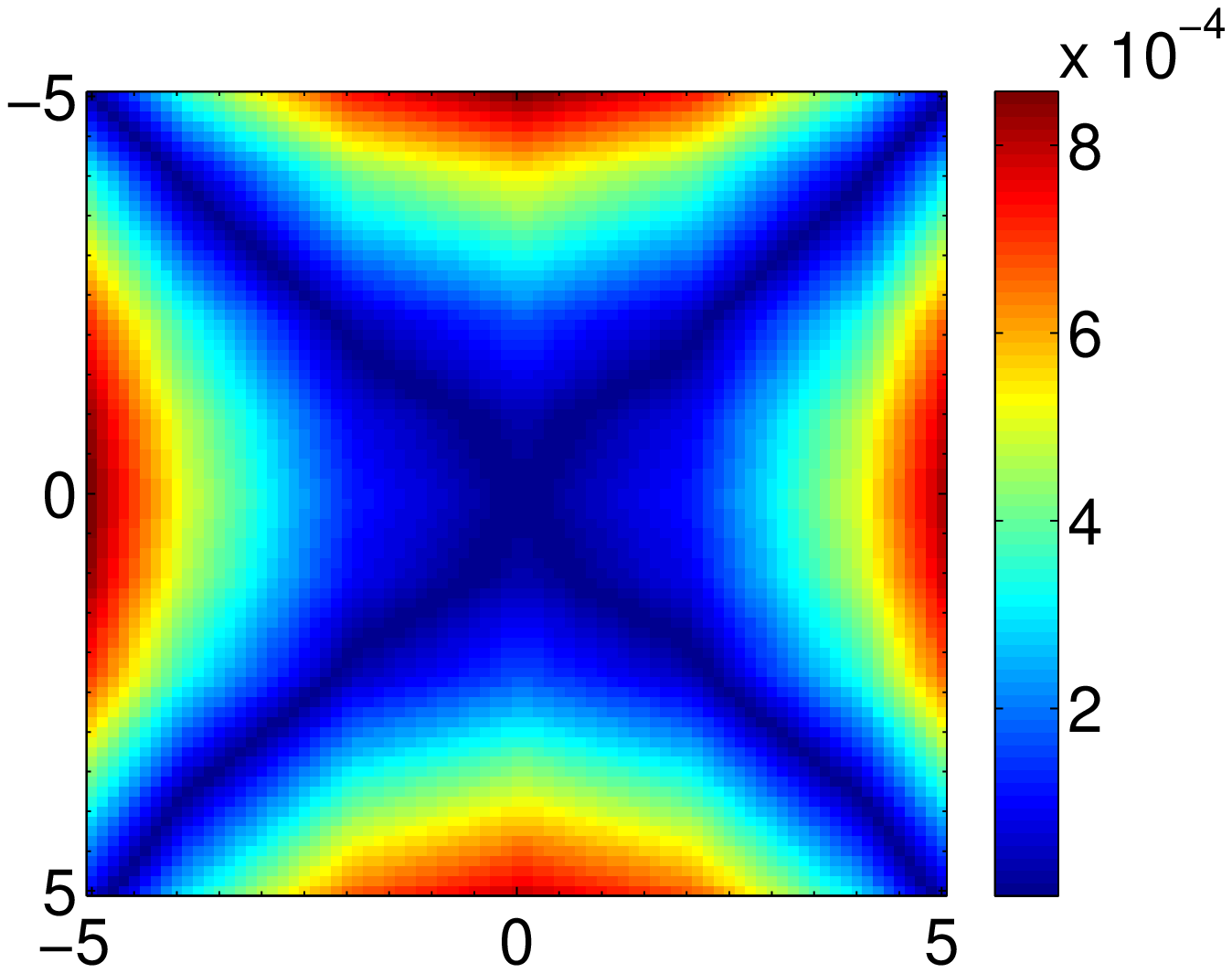}}
\subfigure[Parabolic: Lorentz force in $\text{eV/m}$]
{\includegraphics[width=0.32\textwidth]{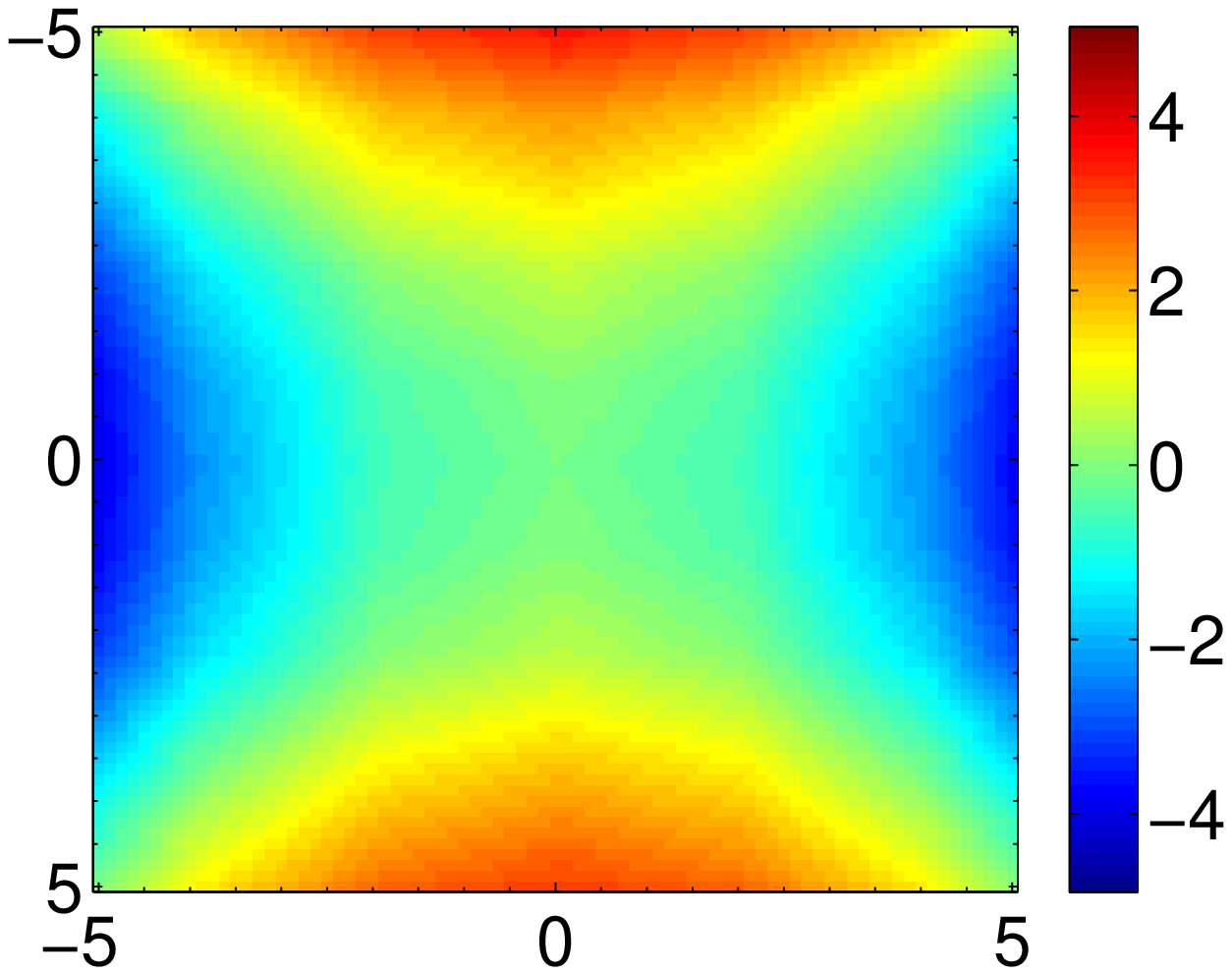}}
\caption{\label{fig:parabolic_vs_flat} Comparison of the field homogeneity of flat-shaped and parabolically-shaped electrodes across the beam extension in the range $\pm x=5$\,mm and   $\pm y=5$\,mm around the beam axis in the center of the RF Wien filter ($z=0$). The top row shows the fields variation of $E_x$ (panel a) and $H_y$ (b), and the Lorentz force (c) for flat-shaped electrodes, and the bottom row the corresponding results for parabolically-shaped electrodes, $E_x$ (d), $H_y$ (e), and the Lorentz force (f). (Note the different in the graphs.)}
\end{figure*}

The advantage of using parabolic electrodes is illustrated in Fig.~\ref{fig:parabolic_vs_flat}, where the relative variation of $E_x$ and $H_y$, \textit{e.g.}, $\left| \frac{|E_x|-\langle E_x \rangle}{|E_x|}\right|$, is shown in the $xy$ plane in  the range $x=\pm 5$\,mm and   $y=\pm 5$\,mm in the center of the RF Wien filter. The results are summarized in Table~\ref{tab:c_variation}, where the relative standard deviation of  panels a), b), d), and e) of Fig.~\ref{fig:parabolic_vs_flat} is listed. With respect to the Lorentz force $\vec F_\text{L}$, parabolically-shaped electrodes show a better homogeneity along the beam trajectory. 

\begin{table}[b]
\renewcommand{\arraystretch}{1.1}
\centering
\caption{Calculated relative standard deviation (RSD) of the electric and magnetic fields in the cases of flat and parabolically-shaped electrodes from Fig.~\ref{fig:parabolic_vs_flat}, quantitatively indicating the achieved field homogeneity.}
\begin{tabular}{ccc}\hline
RSD & flat shape & parabolic shape\\ \hline
$|\frac{\sigma(E_x)}{\langle E_x\rangle}|$ & $4.74 \times 10^{-4}$ & $2.33 \times 10^{-5}$\\
$|\frac{\sigma(H_y)}{\langle H_y\rangle}|$ & $3.17 \times 10^{-5}$ & $ 3.53 \times 10^{-4}$\\\hline
\end{tabular}
\label{tab:c_variation}
\end{table}

The parabolically-shaped electrodes yield an improve\-ment of up to a factor of $20$ in terms of local electric field homogeneity while reducing the local homogeneity of the magnetic field  by a factor of $11$. In total, the Lorentz forces are about a factor $5$ smaller for parabolically-shaped electrodes with respect to flat-shaped electrodes.
 
\section{Beam dynamics simulations \label{sec:beam_dynamics_simulations}}
\subsection{Definition of unwanted field components}
The finite beam size induces non-ideal field components of $\vec{E}$ and $\vec{H}$ in undesired directions. The system is designed such that \textit{unwanted} field components average out at the exit and entry of the RF Wien filter. The unwanted field components are given by
\begin{equation}
\vec{E}_\perp = \left(\begin{array}{ccc}
  0\\
  E_y \\
  E_z\\
 \end{array}\right)  \,, \, \text{and}\,\,
\vec{H}_\perp = \left(\begin{array}{ccc}
  H_x\\
  0 \\
  H_z\\
 \end{array}\right) \label{eq:EH_perp} \,.
\end{equation}
The index $\perp$ is introduced to indicate that the unwanted electric and magnetic field components are perpendicular to the  main field components $E_x$ and $H_y$. 

The particle beam that enters the RF Wien filter has a defined phase-space distribution~\cite{PhysRevSTAB.18.020101}, therefore the particles do not travel along straight lines, parallel to the beam axis. In order to quantify the effect of unwanted field components, a Monte-Carlo simulation based on solving the relativistic equation of motion using the time- and space-dependent fields has been carried out, with subsequent integration of the field components along the trajectories. The electromagnetic simulation yields the complex fields $\tilde{\vec{E}}$ and $\tilde{\vec{H}}$. The real fields $\vec{E} \left(\vec{r},t\right)$ and $\vec{H} \left(\vec{r},t\right)$ are obtained from
\begin{eqnarray}
\vec{E} \left(\vec{r},t\right) & = & \Re\left( \tilde{\vec{E}}e^{i\omega t} \right) \,,\text{and} \nonumber\\
\vec{H} \left(\vec{r},t\right) & = & \Re\left( \tilde{\vec{H}}e^{i\omega t} \right)\,.
\end{eqnarray}
The relativistic equations of motion~\cite{humphries2013charged,wiedemann2015particle},
\begin{eqnarray}
\frac{d\vec{v}}{dt} &=&\frac{q}{m \gamma}\left[ \vec{E}(\vec r,t)+\vec{v}\times \vec{B}(\vec r,t)\right]\nonumber\\
&&-\frac{q}{m \gamma c^2} \vec{v} \left[\vec{v} \cdot \vec{E}(\vec r,t) \right]\,,\text{and} \nonumber\\
\frac{d\vec{r}}{dt} &=&\vec{v}\,,
\end{eqnarray}
were solved in Matlab$\footnote{Mathworks, Inc. Natick, Massachusetts, USA, \url{http://de.mathworks.com}}$ using the field maps imported from the full-wave simulations. The mesh accuracy in the ${xy}$ plane amounts to $\SI{0.1}{mm}$ and $\SI{11}{mm}$ in ${z}$-direction. The phase-space distribution of the beam at the entry of the RF Wien filter is known~\cite{PhysRevSTAB.18.020101}. The phase space is expressed in terms of positions $x$ and $y$, and transverse angles $x'= v_x/v_z$ and $y'= v_y/v_z$. 

\begin{figure*}[htb]
\centering
\subfigure[Low-$\beta$ section ON: $\beta = \SI{0.4}{m}$]{\includegraphics[width=0.40\textwidth]{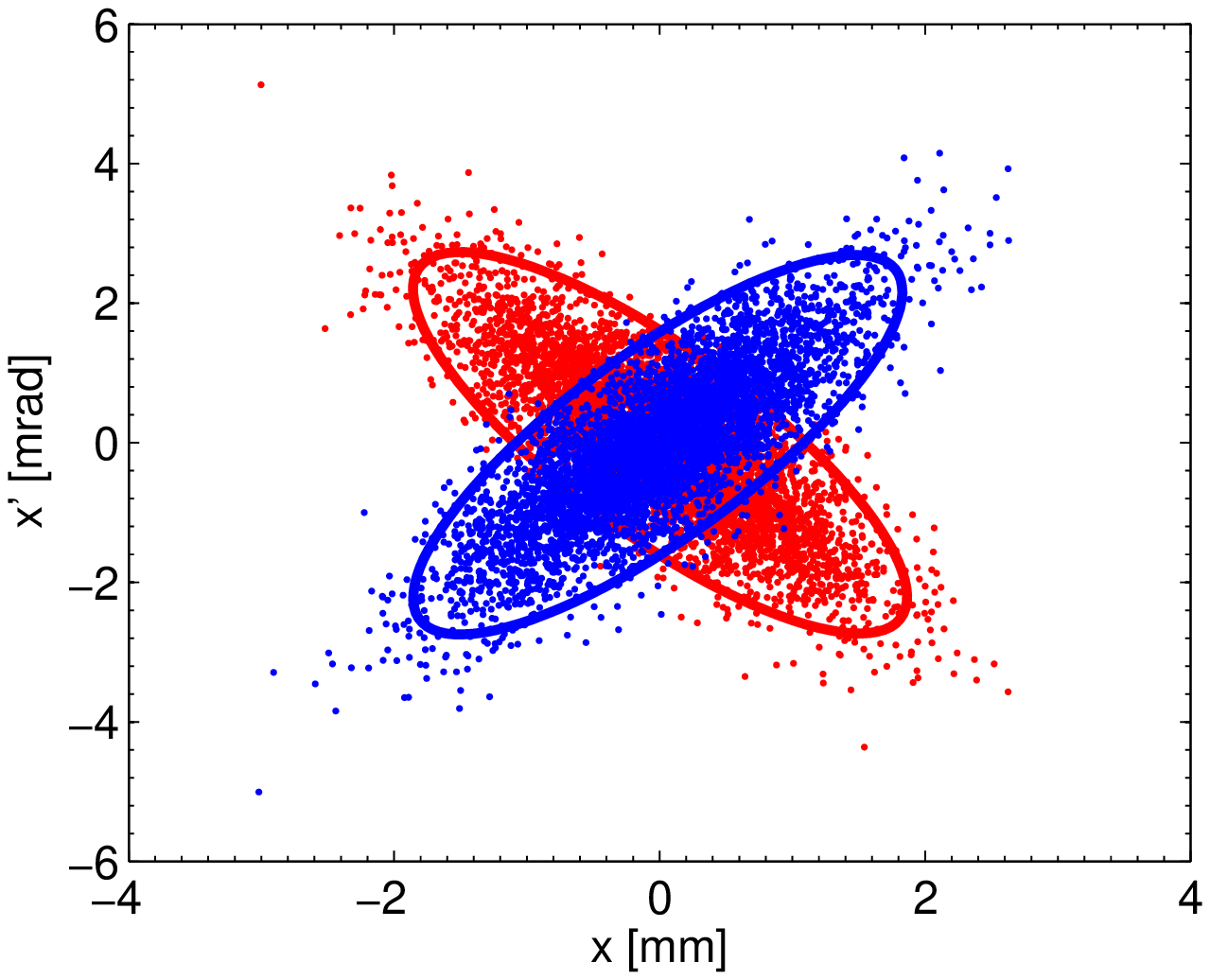}}
\subfigure[Low-$\beta$ section OFF: $\beta = \SI{4}{m}$]{\includegraphics[width=0.40\textwidth]{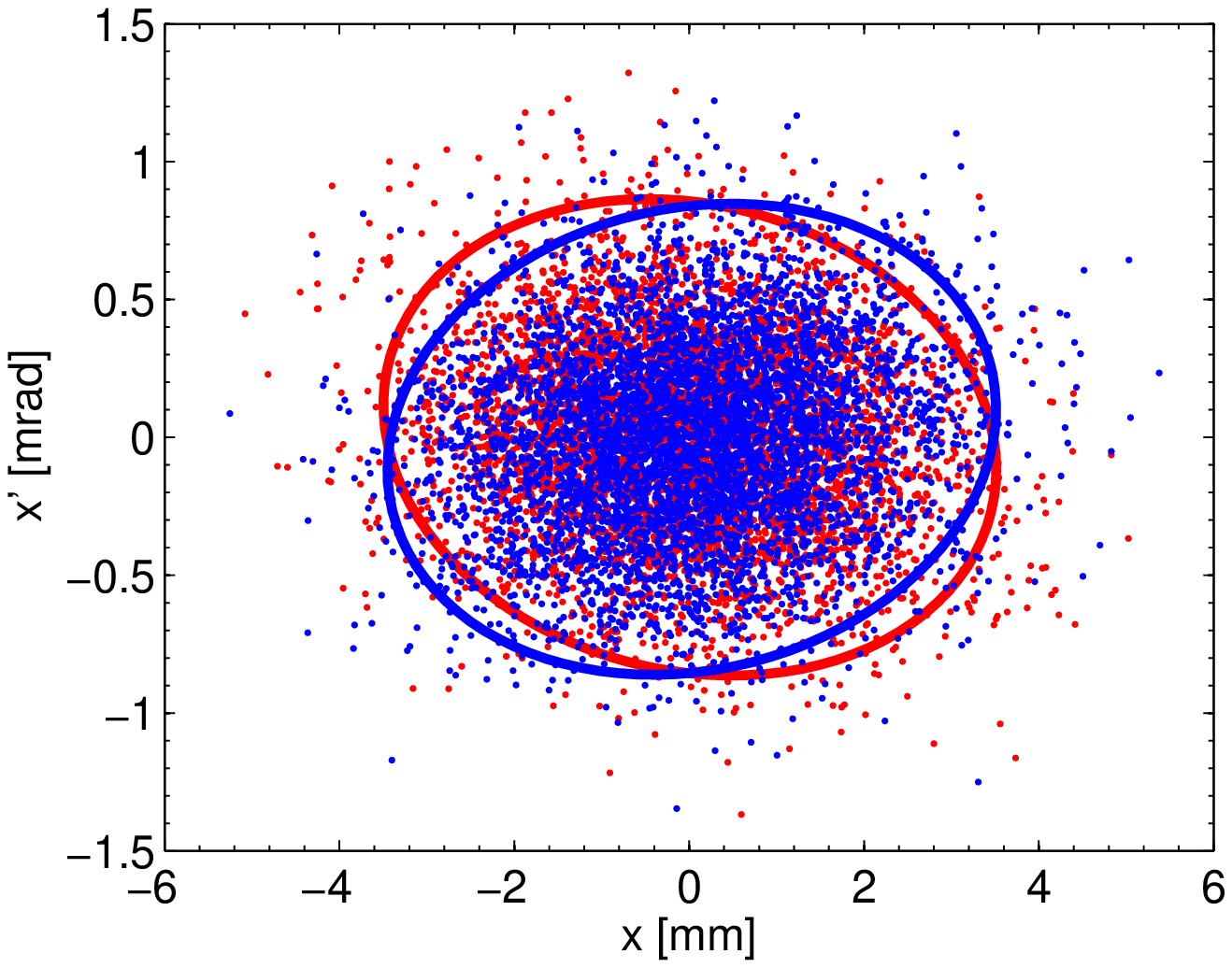}}
\caption{\label{fig:pse_beta} Phase-space distributions $(x,x')$ of a beam with an emittance of $\varepsilon_{x,y}=1$\,$\mu$m entering (red) and exiting (blue) the RF Wien filter with low-$\beta$ section ON:  $\beta = \SI{0.4}{m}$ (panel a) and OFF: $\beta = \SI{4}{m}$ (panel b). The ellipses show the $2 \sigma$ results of a fit with a 2D Gaussian distribution, respectively.}
\end{figure*}

For each phase space, 5000 particles have been simulated with a $2\sigma$ beam emittance of $\varepsilon_{x,y}=1$\,$\mu$m. The initial ${(x,x')}$  phase-space distributions are shown in red in Fig.~\ref{fig:pse_beta}. The blue points represent the phase-space distribution of the particles after passing through the RF Wien filter. The $(y,y')$ distributions are not shown, but are very similar. With low-$\beta$ section ON and OFF, the areas of the initial and final ellipses are the same, and consistent with the behavior of a field free (drift) region. Thus the RF Wien filter does not appear to alter the phase-space distributions of the beam. 

Based on the simulations, the effects of unwanted field components, given in Eq.~(\ref{eq:EH_perp}), are quantified by integration along the particle trajectories, and expressed as ratios with respect to the total field integrals, yielding 
\begin{eqnarray}
f^\text{int}_{E_\perp} &=& \frac{\int |\vec E_\perp|ds}{\int |\vec E| ds} \,\,, \text{and}    \label{eq:e_field_err} \\
f^\text{int}_{H_\perp} &=& \frac{\int |\vec H_\perp|ds}{\int |\vec H| ds}  \label{eq:b_field_err}\,,
\end{eqnarray}  
where $ds$ denotes a differential element of the particle's path, which does not necessarily point along the $z$ axis. 

\subsection{Results of beam dynamics simulations \label{sec:results}}
\begin{figure*}[!]
\centering
\subfigure[Unwanted electric field components $f^\text{int}_{E_\perp}$ from Eq.~(\ref{eq:e_field_err}).]
{\includegraphics[width=0.48\textwidth]{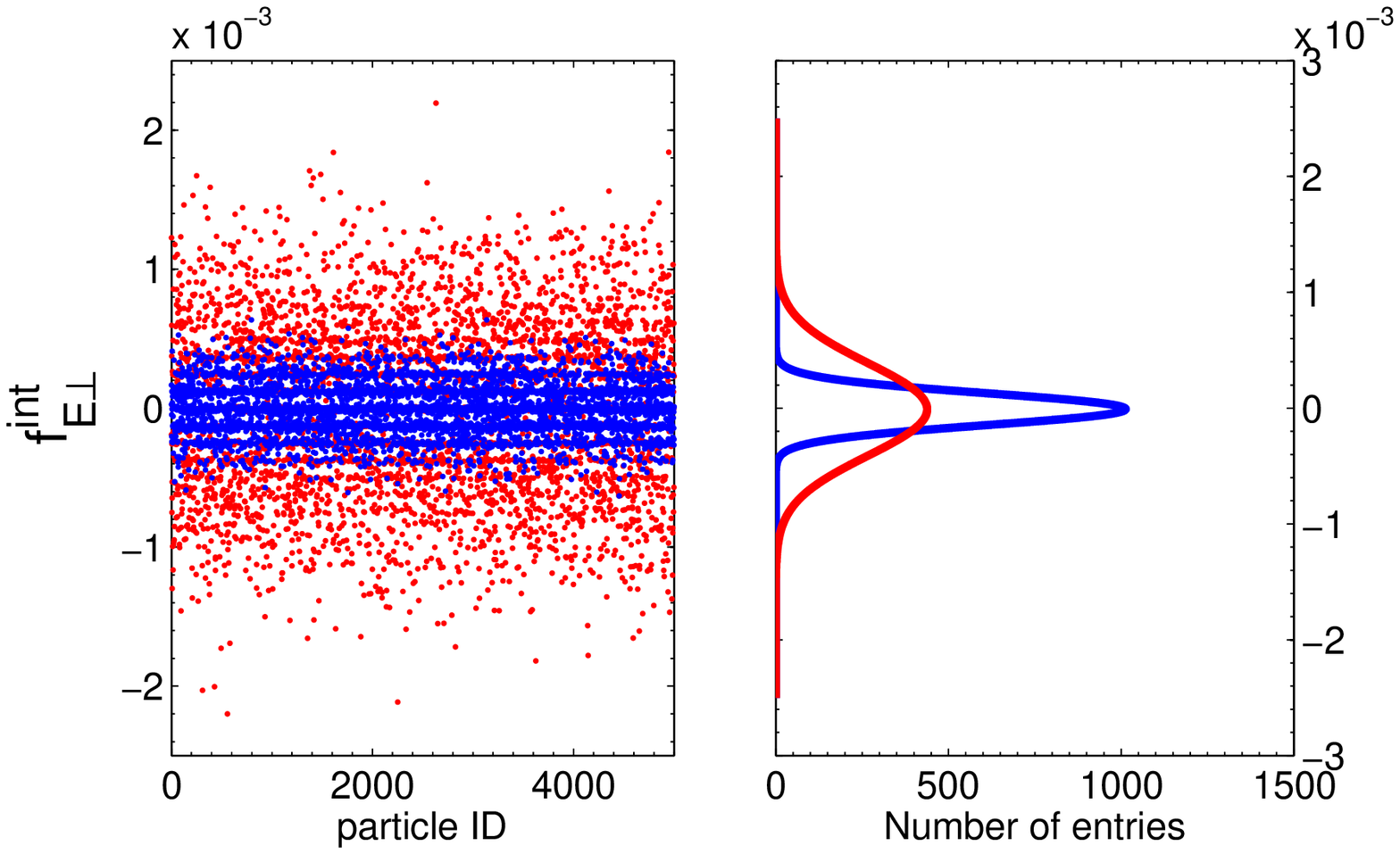}}
\hspace{0.5cm}
\subfigure[Unwanted magnetic field components $f^\text{int}_{H_\perp}$ from Eq.~(\ref{eq:b_field_err}).]  
{\includegraphics[width=0.48\textwidth]{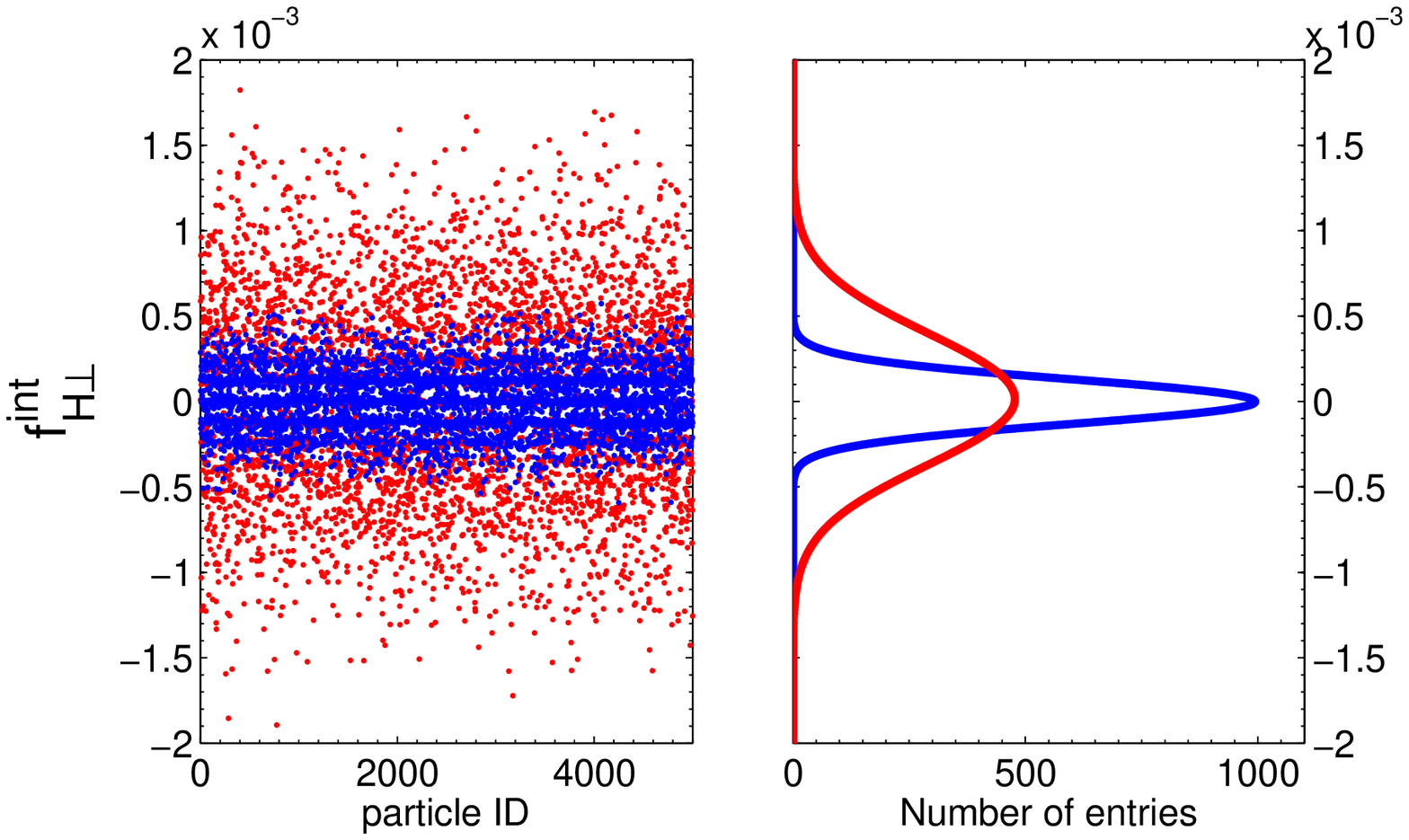}}
\caption{\label{fig:parasitic} Comparison of the unwanted electric (a) and magnetic (b) field components with low-$\beta$ section ON (red) and low-$\beta$ section OFF (blue). The right panels display Gaussian fits to the projections of the simulated distributions, respectively, summarized in Table~\ref{tab:RF-WF-comparison}.}
\end{figure*}

The integrals in Eqs.~(\ref{eq:e_field_err}) and (\ref{eq:b_field_err}) are evaluated for each of the 5000 simulated particles and for the two cases with low-$\beta$ section ON ($\beta=\SI{0.4}{m}$) and OFF ($\beta=\SI{4}{m}$). The results are shown in the left panels of Fig.~\ref{fig:parasitic}, respectively, and summarized in Table~\ref{tab:RF-WF-comparison}. With low-$\beta$ section ON, the beam size is small, and the angular variation is large. As a consequence, the unwanted field components $E_\perp$ and $H_\perp$ do not completely cancel at the edges, leading to a wider distribution. When the low-$\beta$ section is OFF, the beam size is large, but with small angular spread, leading to a better cancellation of the unwanted field components at the edges.

The distributions in Fig.~\ref{fig:parasitic} exhibit a line structure that stems from the finite mesh ($\SI{0.1}{mm}$ in the $xy$ plane, and $\SI{11}{mm}$ along $z$). 
Gaussian fits to the projections of the resulting distributions $f^\text{int}_{E_\perp}$ and $f^\text{int}_{H_\perp}$ are shown in the right panels of Fig.~\ref{fig:parasitic}, the fit results are summarized in Table~\ref{tab:RF-WF-comparison}. For the low-$\beta$ section either switched ON or OFF, the mean values of the integrated unwanted electric and magnetic field components are zero within the errors. The widths of the distributions, however, indicate that with low-$\beta$ section OFF, the unwanted electric and magnetic field components picked up by the beam are smaller by about a factor of two to three compared to when the low-$\beta$ section is ON. These findings should be further investigated using spin and particle tracking of a stored beam. In addition, once the device is installed, the predictions should be verified experimentally.     


\subsection{Comparison of prototype  and waveguide RF Wien filter \label{sec:comparison}}
The waveguide RF Wien filter is compared to the prototype RF Wien filter~\cite{Mey:2015xbq,doi:10.1142/S2010194516600946} (see Section~\ref{sec:introduction}) in terms of the unwanted field components, evaluated according to Eqs.~(\ref{eq:e_field_err}) and (\ref{eq:b_field_err}), listed in Table~\ref{tab:RF-WF-comparison}. With the novel waveguide RF Wien filter, unwanted electric and magnetic field components can be reduced by one to two orders of magnitude. 
\begin{table*}[hbt]
\renewcommand{\arraystretch}{1.1}
\centering
\caption{Comparison of the prototype RF Wien filter~\cite{Mey:2015xbq,doi:10.1142/S2010194516600946} and the waveguide design in terms of  unwanted field components $f^\text{int}_{E_\perp}$ and $f^\text{int}_{H_\perp}$ (see Eqs.~(\ref{eq:e_field_err}) and (\ref{eq:b_field_err})) using the simulated data, shown in Fig.~\ref{fig:parasitic}. 
}
\begin{tabular}{lccccc}\hline
     RF Wien filter    &  & \multicolumn{2}{c}{$f^\text{int}_{E_\perp}$} & \multicolumn{2}{c}{$f^\text{int}_{H_\perp}$}\\
             &                                &  mean $\mu$ & width $\sigma$ & mean $\mu$ & width $\sigma$ \\\hline
\multirow{ 2}{*}{Waveguide} & low-$\beta$ ON  & $(-1.1 \pm 0.8) \times 10^{-5}$ & $5.7 \times 10^{-4}$ & $(1.4 \pm 0.8) \times 10^{-5}$  & $5.5 \times 10^{-4}$\\
                            & low-$\beta$ OFF & $(4.1 \pm 2.5) \times 10^{-6}$ & $1.8 \times 10^{-4}$ & $(0.8 \pm 2.6) \times 10^{-6}$ & $1.8 \times 10^{-4}$\\\hline
Prototype &  & $(1.454 \pm 0.009) \times 10^{-2}$ & $8.9 \times 10^{-3}$ & $(2.927 \pm 0.003) \times 10^{-2}$   & $3.0\times 10^{-3}$\\
\hline
 \end{tabular}
\label{tab:RF-WF-comparison}
\end{table*}

Spin tracking simulation are required to quantify the systematic errors induced by unwanted field components, and by other systematic effects, \textit{e.g.}, positioning errors, and non-vanishing Lorentz forces   in the determination of proton and deuteron EDMs. 

\section{Thermal response of the RF Wien filter during operation \label{sec:thermal}}
Because of the thermal insulation by the vacuum, it was investigated whether even small power losses could lead to a temperature rise during operation. Therefore, for the corresponding thermal simulations using the FEM software ANSYS\footnote{ANSYS, Inc. Canonsburg, USA \url{http://http://www.ansys.com}} only radiative heat exchange between the internal surfaces was considered. The entire power in the copper plates was assumed to be generated at the locations of the feedthroughs. The calculated power loss densities with an input power of $\SI{1}{kW}$ were integrated over the structural elements and are summarized in Table~\ref{tbl_2}. The temperature rise for the copper electrodes is $6$\,K, and $0.5$\,K for the ferrites, respectively. Based on the temperature distribution, the simulated thermal expansion of the copper plates amounts to $48$\,$\mu $m, and to $8.6$\,$\mu$m for the ferrite blocks. These values are small compared to the structure's dimensions, and we do not expect problems with respect to the electromagnetic performance and the suspension of the mechanical structure during operation of the RF Wien filter. 

\begin{table}[htb]
\renewcommand{\arraystretch}{1.1}
  \centering
\caption{\label{tbl_2} Losses per material computed using the electromagnetic solver of CST Microwave Studio at a frequency $f_\text{RF}=\SI{1}{MHz}$.}
  \begin{tabular}{rl}
\hline
   Material        & Thermal losses [W] \\\hline  
   Copper Plates   & 0.0957 \\
   Ferrites        & 0.0734  \\
   Steel           & 0.0255 \\
   Ceramic         & 0.0010 \\\hline
 \end{tabular}
\end{table}

\section{Conclusion and Outlook \label{sec:conclusion}}
The paper presents the electromagnetic design calculations for a novel type of RF Wien filter that shall be used at the COSY storage ring to determine the EDMs of deuterons and protons. The main emphasis of the work was to design a device that exhibits a high level of homogeneity of the electric and magnetic fields. The optimization of the electromagnetic design was performed in close cooperation with the mechanical design, taking into account all the details of the mechanical construction.

A waveguide structure was selected because in this case the orthogonality of $E$ and $B$ fields, required for a Wien filter, can be ensured to a high level of precision. Minimizing the overall Lorentz force, while still providing  sizable electric and magnetic field integrals, leads to parabolically-shaped electrodes, equipped with trapezoid-shaped entrance and exit partitions, surrounded by a closed box of ferrites. Using the above described electrodes, the overall Lorentz force is reduced by about a factor of five, compared to flat-shaped electrodes. 

The RF Wien filter will be installed in a section at COSY, where it is possible to vary the beam size by adjusting the $\beta$ function between about $\beta=0.4$ and $\SI{4}{m}$. Single-pass tracking calculations were performed in order to quantify the effect of unwanted electric and magnetic field components picked up by the beam, realistically distributed in phase space. The calculations indicate that unwanted field components picked up by the beam can be reduced by about a factor of three when the RF Wien filter is operated at $\beta=\SI{4}{m}$. 

Thermal simulations show that the heat load in the different materials of the device, when operated with an input power of $\SI{1}{kW}$ is tolerable. Thus during operation, the mechanical accuracy of the device does not appear to substantially deteriorate the performance.

Forthcoming work will address improvements of the driving circuit, which will allow us to reach larger field values with the same input power. In addition, a novel semi-analytical approach to assess quantitatively the effect of mechanical tolerances on the electromagnetic performance of the RF Wien filter is presently being developed. In addition, spin tracking studies are planned to quantify the impact of the waveguide RF Wien filter on the systematic error of the planned proton and deuteron EDM experiments.

\section*{Acknowledgement}
This work has been performed in the framework of the JEDI (J{\"u}lich Electric Dipole Moment Investigations) collaboration. The authors would like to thank Martin Gaisser, Volker Hejny, Kirill Grigoryev, and J{\"o}rg Pretz for valuable comments on the manuscript.

\section*{References}

\bibliography{dBase}

\end{document}